\documentstyle[12pt]{article}
\def\be{\begin{equation}}
\def\ee{\end{equation}}
\def\bea{\begin{eqnarray}}
\def\eea{\end{eqnarray}}

\begin{document}

\begin{center}
{\Large{\bf Fermions in the Presence of the Antisymmetric Fields}}

\vskip .5cm
{\large Davoud Kamani}
\vskip .1cm
 {\it Faculty of Physics, Amirkabir University of Technology (Tehran Polytechnic) 
\\  P.O.Box: 15875-4413, Tehran, Iran}\\
{\sl e-mail: kamani@cic.aut.ac.ir}
\\
\end{center}

\begin{abstract}

In this manuscript we study the Dirac action in the presence of the Ramond-Ramond (R-R)
potentials as gauge fields. Therefore, for the R-R field $A_{\mu_1...\mu_{p+1}}$,
we identify the corresponding fermion with an extended $p$-dimensional
object, which we call it F$p$-brane.
Conservation of the tensor currents, associated to these fermionic
branes, imposes an external tensor
current. This external current enables us to study the R-R fields and their Hodge
dual fields as independent degrees of freedom.
We observe that an F$p$-brane should live with its dual brane,
$i.e.$ F$(d-p-2)$-brane. The gauge symmetry and some other properties of
a system of an F$p$-brane and its dual object will be discussed.

\end{abstract}

{\it PACS}: 11.25.-w

{\it Keywords}: Dirac action; R-R fields; Branes.

\vskip .5cm
\newpage

\section{Introduction}

Supergravity theories in diverse dimensions admit a variety of
$p$-branes. In the context of the effective $d=10$ or $d=11$
dimensional supergravity theories, a $p$-brane is a
$p$-dimensional extended source for a $(p+2)$-form field strength
$F_{p+2}$. This field satisfies the equation
\bea
\nabla_\mu F^{\mu\mu_1...\mu_{p+1}} = j^{\mu_1...\mu_{p+1}},
\eea
where $j^{\mu_1...\mu_{p+1}}$ is a $(p+1)$-form current. The dual field
$F'$ also satisfies the equation
\bea
\nabla_\mu F'^{\mu\mu_1...\mu_{p'+1}} = j'^{\mu_1...\mu_{p'+1}},
\eea
in which $j'^{\mu_1...\mu_{p'+1}}$ is $(p'+1)$-form current with $p' =
d-p-4$ \cite{1,2,3}.

From the other side, central extension of the superalgebras induces fermionic charges
on the branes \cite{4,5}. By some assumptions the fermionic charge of D-string is
interpreted as a source for the dilaton field \cite{6}. Similarly, modification of the
super-AdS algebras imposes fermionic charges for the branes \cite{7}.
Note that gauging an scalar field theory by
form fields also is possible, $e.g.$ see \cite{8}.

We construct a model which incorporates fermionic brane charge, whose existence
has been anticipated in the Refs. \cite{4,5,6,7}. Therefore,
combining the R-R fields and the fermionic charges as physical bases of our model, we
introduce the gauged Dirac action by the R-R fields. This is analog of QED. That is,
extended fermions are sources for the R-R fields. Assume there is a fermionic field
$\psi_p$ corresponding to a $p$-dimensional brane. A tensor current due to this
fermion appears as source of the R-R field $A_{\mu_1...\mu_{p+1}}$. In other words, this
R-R field can be emitted by $\psi_p$. We call this $p$-dimensional
fermionic object as ``F$p$-brane'',
where ``F'' refers to the fermionic properties. For this system the gauge symmetry is obvious.

Generally, an F$p$-brane can be viewed as a new object which is
different from a D$p$-brane. In contrast to the D-branes, the
Hodge dual of an F$p$-brane is $(d-p-2)$-dimensional object. We
shall observe that an F$p$-brane should live with its dual brane.
This system under an external current is formed. The external current
generalizes the equations (1) and (2). Therefore, the R-R fields and
their Hodge dual fields appear as independent degrees of freedom.

This paper is organized as follows. In section 2, the gauged
Dirac action by an R-R field will be studied. In section 3, the
Hodge dual of the F$p$-brane, $i.e.$ F$(d-p-2)$-brane will be
introduced. In section 4, a system of an F$p$-brane with
F$(d-p-2)$-brane in the presence of an external current will be
analyzed.
\section{The gauged action}

For the Dirac action in the $d$-dimensional flat spacetime, which
is gauged by the R-R field $A_{\mu_1...\mu_{p+1}}$, we introduce the
following action
\bea
S= \int d^d X \bigg{(} {\bar
\psi}_p(i\Gamma^\mu D_\mu -m )\psi_p -\frac{1}{2(p+2)!}
F_{\mu_1...\mu_{p+2}}F^{\mu_1...\mu_{p+2}} \bigg{)},
\eea
where $\psi_p$ is a Dirac field, $D_\mu$ is an appropriate
operator, $m$ is the mass of the F$p$-brane and $F_{p+2}$ is
field strength of $A_{p+1}$, $i.e.$ $F_{p+2}=dA_{p+1}$. The Dirac
matrices are $\{\Gamma^\mu\}$.
The spacetime metric also is considered as
$\eta_{\mu\nu}={\rm diag}(-1,1,...1)$.
Since the pre-factor of the integral will not appear in our calculations,
we put it away.

We shall see that the gauge symmetry of the action (3) admits the
following form for the operator $D_\mu$, 
\bea 
D_\mu = \partial_\mu + iq_p
\Gamma^{\mu_1...\mu_p}A_{\mu\mu_1...\mu_p},
\eea 
where $\Gamma^{\mu_1...\mu_p}=
\Gamma^{[\mu_1}\Gamma^{\mu_2}...\Gamma^{\mu_p]}$ is totally antisymmetric.
The constant factor $q_p$ is the R-R charge corresponding to the
F$p$-brane. Thus, the action (3) describes the fermionic field
$\psi_p$ and the R-R field $A_{\mu_1...\mu_{p+1}}$, which
interact with each other. The interaction term of $\psi_p$ with
$A_{\mu_1...\mu_{p+1}}$ reveals that $\psi_p$ can emit (absorb)
this R-R field. In addition, $A_{\mu_1...\mu_{p+1}}$ can also split
to $\psi_p$ and ${\bar \psi_p}$. Since this R-R field has a
$p$-dimensional brane source, we can say that F$p$-brane is an
extended fermionic object.

Note that for simplicity we consider flat spacetime. It is
straightforward to write the action $S$ in the curved background.
\subsection{Field equations}

Vanishing the variation of the action (3) gives the following equations of
motion
\bea
(i\Gamma^\mu D_\mu - m)\psi_p =0, 
\eea 
\bea
\partial_\nu F^{\nu\mu\mu_1...\mu_p} - q_p j_p^{\mu \mu_1...\mu_p}=0,
\eea
where the $(p+1)$-form current $j_p^{\mu \mu_1...\mu_p}$ has the
definition 
\bea
j_p^{\mu \mu_1...\mu_p} = {\bar \psi}_p \Gamma^{\mu\mu_1...\mu_p} \psi_p.
\eea
This is an extended current corresponding to the F$p$-brane. The
equation (6) emphasizes that this current reveals a fermionic
source for the R-R field $A_{\mu\mu_1...\mu_p}$.

Derivative of the equation (6) gives the conservation law
\bea
\partial_\mu j_p^{\mu\mu_1...\mu_p}=0.
\eea
Note that effect of any $\partial_{\mu_i}$ with $\mu_i \in
\{\mu,\mu_1,...,\mu_p\}$ leads to the equation (8). In fact, this equation
puts some conditions on the fermionic field $\psi_p$. 
Since the indices are chosen from the set
$\{0,1,...d-1\}$, the equation (8) (for $p\geq1$) puts
$\frac{d!}{p!(d-p)!}$ conditions on $\psi_p$. By modifying the
action we shall remove these conditions.
\subsection{Gauge symmetry}

Now we discuss the gauge symmetry of the action (3).
For this, consider the gauge transformations
\bea
&~& A_{p+1} \longrightarrow A_{p+1} + d \Lambda,
\nonumber\\
&~& \psi_p \longrightarrow e^{i\alpha.\Lambda}\psi_p ,
\eea
where $\Lambda(X)$ and $\alpha(X)$ are local $p$-forms.
The dot product $\alpha \cdot \Lambda$ means
$\alpha \cdot \Lambda \equiv \alpha^{\mu_1 ...\mu_p}\Lambda_{\mu_1 ...\mu_p}$.
If $D_\mu$ really is covariant derivative, $D_\mu \psi_p$ should have gauge transformation
like the field $\psi_p$. Therefore, having the transformation
\bea
D_\mu \psi_p \longrightarrow e^{i\alpha.\Lambda} D_\mu \psi_p,
\eea
leads to the following equation
\bea
\partial_\mu (\alpha.\Lambda) =
-(p+1)q_p\Gamma^{\mu_1...\mu_p}\partial_{[\mu}\Lambda_{\mu_1...\mu_p]}.
\eea
According to this equation, the action $S$ is invariant
under the transformations (9).

The equation (11) gives $d$ relations between the components of
$\alpha$, while $\Lambda$ remains arbitrary. Thus, up to these
relations, the functions $\{\alpha^{\mu_1...\mu_p}(X)\}$ also are
arbitrary. As an example, for F0-brane, the solution of (11) is
\bea 
\alpha (X)= -q_0 + \frac{C}{\Lambda (X)}, 
\eea 
where $C$ is an arbitrary constant.
\section{Action of the dual fields}

Now we introduce the Hodge duals of the fields $\psi_p$ and
$A_{p+1}$. Let denote them by $\psi'_{p'}$ and $A'_{p'+1}$. The R-R field
$A'_{\mu_1...\mu_{p'+1}}$ is given by
\bea
A'_{\mu_1...\mu_{p'+1}} = \frac{1}{(p+1)!}
\epsilon_{\mu_1...\mu_{p'+1}}\;^{\nu_1...\nu_{p+1}}
A_{\nu_1...\nu_{p+1}}.
\eea
In this section we do not use the
equation (13). That is, for the next purposes we study the
behavior of $A'_{p'+1}$ as an independent degree of freedom. This
implies that the action should have the term $|dA'_{p'+1}|^2$. In
addition, $A'_{\mu_1...\mu_{p'+1}}$ appears in the covariant
derivative. Thus, the action of the dual variables also has the
structure of $S$, $i.e.$,
\bea
S' = \int d^dX\bigg{(} {\bar
\psi'}_{p'}(i\Gamma^\mu D'_\mu - m')\psi'_{p'}
-\frac{1}{2(p'+2)!} F'_{\mu_1...\mu_{p'+2}}F'^{\mu_1...\mu_{p'+2}}
\bigg{)}.
\eea
The form $F'_{p'+2}$ is field strength of $A'_{p'+1}$,
\bea
F'_{p'+2}=dA'_{p'+1} .
\eea
The covariant derivative $D'_\mu$ is given by (4)
with $A'_{\mu\mu_1...\mu_{p'}}$ instead of $A_{\mu\mu_1...\mu_p}$
and also $p'$ instead of $p$, where $p'=d-p-2$ is the dimension
of an F$p'$-brane. This brane is source of the R-R field
$A'_{p'+1}$. It has the R-R charge $q_{p'}$ and the mass $m'$.
The corresponding fermionic field is $\psi'_{p'}$. This field is
defined as the dual of $\psi_p$. In fact, through the
equation of motion $\psi'_{p'}$ depends on the Hodge dual of $A_{p+1}$. 

Note that in the action $S'$ we applied the field strength
$F'_{p'+2}=d*A_{p+1}$.
This is different from the D-brane case that $F'_{p'+2}= *dA_{p+1}$.
Therefore, for a D$p$-brane the Hodge dual is
D$(d-p-4)$-brane, while in our model the dual of an F$p$-brane is F$(d-p-2)$-brane.
In other words, an F$p$-brane is different from a D$p$-brane.

The action $S'$ under the gauge transformations
\bea
&~& A'_{p'+1} \longrightarrow A'_{p'+1} + d\Lambda' ,
\nonumber\\
&~& \psi'_{p'} \longrightarrow e^{i\alpha'.\Lambda'} \psi'_{p'} ,
\eea
is invariant. This invariance implies the following relation
between the components of $\alpha'^{\mu_1...\mu_{p'}}$,
\bea
\partial_\mu(\alpha' . \Lambda')=-(p'+1)q_{p'} \Gamma^{\mu_1...\mu_{p'}}
\partial_{[\mu}\Lambda'_{\mu_1...\mu_{p'}]} ,
\eea

According to the equations of motion, extracted from the action
$S'$, the fermionic field $\psi'_{p'}$ should satisfy the
condition
\bea
\partial_\mu {j'}_{p'}^{\mu\mu_1...\mu_{p'}}=0.
\eea
In deed, this is conservation law for the current 
${j'}_{p'}^{\mu\mu_1...\mu_{p'}}$. This current originates
from the dual fermions. By combining the actions $S$ and $S'$ the
conditions (8) and (18) will be removed.
\section{A system of F$p$-brane and F$(d-p-2)$-brane}

Now we proceed to study a system of an F$p$-brane with an F$p'$-brane
where $p'=d-p-2$.
That is, the fields $A_{\mu_1...\mu_{p+1}}$ and
$A'_{\mu_1...\mu_{p'+1}}$ (and also $\psi_p$
and $\psi'_{p'}$) appear as independent degrees of freedom. This
can be done by combining the actions $S$ and $S'$. In addition,
we consider the equation (13) as a constraint. Adding all these
together, we obtain the action
\bea
I=\int d^dX \bigg{[} {\bar
\psi_p}(i\Gamma^\mu D_\mu -m)\psi_p -\frac{1}{2(p+2)!}
F_{\mu_1...\mu_{p+2}}F^{\mu_1...\mu_{p+2}}
\nonumber\\
+{\bar \psi'_{p'}}(i\Gamma^\mu D'_\mu -m')\psi'_{p'}
-\frac{1}{2({p'}+2)!}
F'_{\mu_1...\mu_{{p'}+2}}F'^{\mu_1...\mu_{{p'}+2}}
\nonumber\\
-(p'+1)! J^{\mu_1...\mu_{p+1}} \bigg{(} A_{\mu_1...\mu_{p+1}}
-\frac{\eta}{(p'+1)!}
\epsilon_{\mu_1...\mu_{p+1}}\;^{\nu_1...\nu_{p'+1}}
A'_{\nu_1...\nu_{p'+1}} \bigg{)} \bigg{]}.
\eea
The tensor field
$J^{\mu_1...\mu_{p+1}}$ with the conventional factor $(p'+1)!$
is Lagrang multiplier. The factor $\eta$ is the effect of double
Hodge duality, $i.e.$ on a $(p+1)$-form it is $\eta = ** =
(-1)^{p(d-p)+d}$.

The equations of motion, extracted from the action (19), are as
in the following
\bea
&~&(i\Gamma^\mu D_\mu -m)\psi_p=0 ,
\nonumber\\
&~& (i\Gamma^\mu D'_\mu -m')\psi'_{p'}=0 ,
\nonumber\\
&~& A_{\mu_1...\mu_{p+1}} -\frac{\eta}{(p'+1)!}
\epsilon_{\mu_1...\mu_{p+1}}\;^{\nu_1...\nu_{p'+1}}
A'_{\nu_1...\nu_{p'+1}} =0 ,
\nonumber\\
&~& \partial_\nu F^{\nu\mu\mu_1...\mu_p} - q_p 
j_p^{\mu\mu_1...\mu_p} = (p'+1)! J^{\mu\mu_1...\mu_p} ,
\nonumber\\
&~& \partial_\nu F'^{\nu\mu\mu_1...\mu_{p'}} - q_{p'}
{j'}_{p'}^{\mu\mu_1...\mu_{p'}} = (p+1)!
J'^{\mu\mu_1...\mu_{p'}}, 
\eea 
where the $(p'+1)$-form
$J'_{p'+1}$ is Hodge dual of the $(p+1)$-form $J_{p+1}$, \bea
J'^{\mu\mu_1...\mu_{p'}} =
\frac{1}{(p+1)!}\epsilon^{\mu\mu_1...\mu_{p'}}
\;_{\nu_1...\nu_{p+1}}J^{\nu_1...\nu_{p+1}} .
\eea

In the fourth and fifth equations of (20) the tensor fields
$J^{\mu\mu_1...\mu_p}$ and $J'^{\mu\mu_1...\mu_{p'}}$ act as the
external currents. We can also see this from the
action (19). Therefore, the third line of (19) can be written as
\bea
{\cal{L}}_c = -((p'+1)!J . A + (p+1)!J' . A' ).
\eea
This Lagrangian density reveals that $J^{\mu\mu_1...\mu_p}$ and
$J'^{\mu\mu_1...\mu_{p'}}$ are external sources corresponding to
the fields $A_{\mu\mu_1...\mu_p}$ and $A'_{\mu\mu_1...\mu_{p'}}$,
respectively.
In other words, the constraint-terms in the action find the
feature of the source-terms.

The fourth and fifth equations of (20) lead to the following
conservation laws
\bea
&~& \partial_\mu (q_p j_p^{\mu\mu_1...\mu_p}
+(p'+1)! J^{\mu\mu_1...\mu_p})=0, 
\nonumber\\
&~& \partial_\mu (q_{p'} {j'}_{p'}^{\mu\mu_1...\mu_{p'}} +(p+1)!
J'^{\mu\mu_1...\mu_{p'}})=0.
\eea
The first equation of (23) implies that
the combination of the external current and the current due to
the fermionic field $\psi_p$ is conserved. By the second
equation, this also holds for the dual case. These equations also
removed the conditions (8) and (18) from the fields $\psi_p$ and
$\psi'_{p'}$.

In fact, the conservation laws (23) impose the external current
$J^{\mu\mu_1...\mu_p}$. This current interacts with the R-R
fields, emitted by the F$p$-brane and the F$(d-p-2)$-brane.
Therefore, the action (19) implies that in the presence of the
external current a system of the F$p$-brane with the
F$(d-p-2)$-brane is formed.
\subsection{Symmetries of the action $I$}

Under the gauge transformations (9) and (16) (with the equations (11) and (17) for the
elements of $\alpha$ and $\alpha'$) the action $I$ is invariant if the forms
$\Lambda$ and $\Lambda'$ have the relation
\bea
d\Lambda' = *d\Lambda .
\eea
Note that the external current $J_{p+1}$, and hence $J'_{p'+1}$, should have trivial
transformations.

Another symmetry is as follows. Under the exchanges of the variables with their dual
variables, $i.e.$,
\bea
&~& A_{p+1} \longleftrightarrow A'_{p'+1} ,
\nonumber\\
&~& \psi_p \longleftrightarrow \psi'_{p'} ,
\nonumber\\
&~& J_{p+1} \longleftrightarrow J'_{p'+1} ,
\eea
and also the exchanges $m \longleftrightarrow m'$ and $p \longleftrightarrow p'$, the action
(19) is symmetric. In fact, the first exchange of (25) gives $q_p \longleftrightarrow q_{p'}$.
The invariance of the source terms of (19), by the equation (22), is more obvious.
\section{Conclusions}

Gauging the Dirac action by the R-R fields produces the fermionic
branes. That is, coupling of a fermionic field with an R-R field leads
to an effective tensor current as a source for the R-R field. Thus, this
fermion describes an extended object, $i.e.$ F$p$-brane.
The gauged action has the gauge symmetry, as expected.

For an F$p$-brane there is a dual brane, $i.e.$ F$(d-p-2)$-brane.
The associated gauge fields of these
branes are Hodge dual of each other. They have different fermionic fields.
The dual theory has its own gauge symmetry.
In general, the F-branes are different from the D-branes.

Conservation of the tensor currents puts some conditions on the above fermionic fields.
In other words, only under these conditions a single F$p$-brane can exist. Removing these
conditions imposes a system of an F$p$-brane with an
F$(d-p-2)$-brane in the presence of an external
current. Therefore, the source of an R-R field is combination of the internal and
the external currents. In addition to the gauge symmetry,
this combined system under the exchange of
the variables with their dual variables is invariant.


\begin{thebibliography}{99}

\bibitem{1}
A. Strominger, Nucl. Phys. {\bf B343} (1990) 167.
\bibitem{2}
M.J. Duff, R.R. Khuri and J.X. Lu, Phys. Rep. {\bf 259} (1995) 213, hep-th/9412184.
\bibitem{3}
C.M. Hull, Nucl. Phys. {\bf B509} (1998) 216, hep-th/9705162.
\bibitem{4}
M.B. Green, Phys. Lett. {\bf B223} (1989) 157.
\bibitem{5}
E. Bergshoeff and E. Sezgin, Phys. Lett. {\bf B354} (1995) 256, hep-th/9504140;
E. Sezgin, Phys. Lett. {\bf B392} (1997) 323, hep-th/9609086.
\bibitem{6}
M. Hatsuda and M. Sakaguchi, Nucl. Phys. {\bf B577} (2000) 183, hep-th/0001214.
\bibitem{7}
K. Peeters and M. Zamaklar, Phys. Rev. {\bf D69} (2004) 066009,
hep-th/0311110.
\bibitem{8}
G. Dvali, ``Three-Form Gauging of axion Symmetries and Gravity'',
hep-th/0507215.


\end{thebibliography}
\end{document}